# Cyberthreat Intelligence:
# Challenges and Opportunities

Mauro Conti, Ali Dehghantanha, and Tooska Dargahi

**Abstract** The ever increasing number of cyber attacks requires the cyber security and forensic specialists to detect, analyze and defend against the cyberthreats in almost real-time. In practice, timely dealing with such a large number of attacks is not possible without deeply perusing the attack features and taking corresponding intelligent defensive actions – this in essence defines cyberthreat intelligence notion. However, such an intelligence would not be possible without the aid of artificial intelligence, machine learning and advanced data mining techniques to collect, analyse, and interpret cyber attack evidences. In this introductory chapter we first discuss the notion of cyberthreat intelligence and its main challenges and opportunities, and then briefly introduce the chapters of the book which either address the identified challenges or present opportunistic solutions to provide threat intelligence.

## 1 Introduction

In the era of digital information technology and connected devices, the most challenging issue is ensuring the security and privacy of the individuals' and organizations' data. During the recent years, there has been a significant increase in the number and variety of cyber attacks and malware samples which make it extremely difficult for security analysts and forensic investigators to detect and defend against such security attacks. In order to cope

Mauro Conti
University of Padua, Italy e-mail: `conti@math.unipd.it`

Ali Dehghantanha
Department of Computer Science, University of Salford, UK, e-mail: `A.Dehghantanha@salford.ac.uk`

Tooska Dargahi
CNIT - University of Rome Tor Vergata, Italy, e-mail: `tooska.dargahi@uniroma2.it`

with this problem, researchers introduced the notion of *"Threat Intelligence"*, which refers to *"the set of data collected, assessed and applied regarding security threats, threat actors, exploits, malware, vulnerabilities and compromise indicators"* (verbatim quoting [9]). In fact, cyberthreat intelligence (CTI) emerged in order to help security practitioners in recognizing the indicators of cyber attacks, extracting information about the attack methods, and consequently responding to the attack accurately and in a timely manner. Here an important challenge would be: How to provide such an intelligence? When a significant amount of data is collected from or generated by different security monitoring solutions, intelligent big-data analytical techniques are necessary to mine, interpret and extract knowledge out of the collected data. In this regard, several concerns come along and introduce new challenges to the filed, which we discuss in the following.

## Cyberthreat Intelligence Challenges

As a matter of fact, Cybercriminals adopt several methods to attack a victim in order to (i) steal their sensitive personal information (e.g., financial information); or (ii) access and take control of the victim's machine to perform further malicious activities, such as delivering malware (in case of Botnet), locking/encrypting victim machine (in case of Ransomware). Though, different cyber attacks seem to follow different methods of infection, in essence they have more or less similar life cycle: starting from victim reconnaissance to performing malicious activities on the victim machine/network.

### Attack Vector Reconnaissance

An important challenge in defending against cyber attacks, is recognizing the point of attacks and the system vulnerabilities that could be exploited by the cybercriminals. Along with the common methods that have always been used to deceive victims (e.g., phishing [11]) in performing the actions that the attackers desire, during the recent years, attackers have used smarter and more innovative methods for attacking victims. These methods are ranging from delivering a malicious software (malware) in an unexpected format (e.g., Word documents or PDF files) to the victim machine [3], to exploiting 0-day vulnerabilities[1], and trespassing anonymous communications in order to contact threat actors [4]. Some examples of such advanced attacks are the new families of Ransomware that have worm-like behaviours, which have infected tens of hundreds of individuals, organizations and critical systems. These advancements in attack methods make the recognition of the attacker and attack's point of arrival an extremely challenging issue.

---

[1] A application vulnerability that is undisclosed and could be exploited by the attackers to access the victim's machine [7]

*Attack Indicator Reconnaissance*

Another important issue regarding the emerging cyber attacks is the fact that cybercriminals use advanced anti-forensics and evasion methods in their malicious code, which makes the usual security assessment techniques, e.g., CVSS (Common Vulnerability Scoring System), or static malware and traffic analysis less efficient [8, 10]. Moreover, the new networking paradigms, such as software-defined networking (SDN), Internet of Things (IoT), and cloud computing, and their widely adoption by organizations (e.g., using cloud resources for their big-data storage and processing) call for modern techniques in forensic investigation of exchanged and stored data [5, 12].

## Cyberthreat Intelligence Opportunities

In order to address the challenges explained in the previous section, the emerging field of cyberthreat intelligence considers the application of artificial intelligence and machine learning techniques to perceive, reason, learn and act intelligently against advanced cyber attacks. During the recent years, researchers have taken different artificial intelligence techniques into consideration in order to provide the security professionals with a means of recognizing cyberthreat indicators. In particular, there is an increasing trend in the usage of Machine Learning (ML) and data mining techniques due to their proved efficiency in malware analysis (in both static and dynamic analysis), as well as network anomaly detection [1, 2, 10]. Along with the methods that the cyber defenders could use in order to prevent or detect cyber attacks, there are other mechanisms that could be adopted in order to deceive the attackers, such as using honeypots. In such mechanisms, security specialists provide fake information or resources that seem to be legitimate to attract attackers, while at the same time they monitor the attackers' activities and proactively detect the attack [6]. Totally, a combination of these methods would be required to provide up-to-date information for security practitioners and analysts.

## 2 A brief review of book chapters

This book provides an up-to-date and advanced knowledge, from both academia and industry, in cyberthreat intelligence. In particular, in this book we provide wider knowledge of the field with specific focus on the cyber attack methods and processes, as well as combination of tools and techniques to perceive, reason, learn and act on a wide range of data collected from different cyber security and forensics solutions.

The remainder of the book is structured as follows. The first six chapters discuss, in details, how the adoption of artificial intelligence would advance the cyberthreat intelligence in several contexts, i.e., in static malware analysis (Chapter 2), network anomaly detection (Chapter 3 and 4), Ransomware detection (Chapter 5 and 6), and Botnet detection (Chapter 7). The next chapter (Chapter 8) presents an investigative analysis of mobile-specific phishing methods and the results of a case study by PayPal. Chapter 9 reviews several methods of malicious payload delivery through PDF files and provides a taxonomy of malicious PDF detection methods. Chapter 10 presents a traffic fingerprinting algorithm for Darknet threat intelligence, which in essence serves as an adaptive traffic association and BGP interception algorithm against Tor networks. Chapter 11 investigates the effectiveness of existing CVSS evaluation results and proposes a model for CVSS analysis suggesting improvements to the calculation of CVSS scores (to be used for Android and iOS applications). Chapter 12 studies the attributes of an effective fake content generator to be used to deceive cyber attackers, and presents an implementation design for an efficient honeypot proxy framework. Chapter 13 discusses possible attacks to the memory data of VMs (Virtual Machines) during live migration in the cloud environment and proposes a secure live VM migration. Chapter 14 introduces a forensic investigation framework for SDN, validating its efficiency considering two use-case scenarios. Finally, the last two chapters warp up the book by assessing and reviewing the state-of-the-art in mobile forensics (Chapter 15) and cloud forensics (Chapter 16).

## Acknowledgement

We would like to sincerely thank all the authors and reviewers, as well as Springer editorial office for their effort towards the success of this book.

## References

1. Ding, Q., Li, Z., Haeri, S., Trajković, L.: Application of machine learning techniques to detecting anomalies in communication networks: Classification algorithms. In: M. Conti, A. Dehghantanha, T. Dargahi (eds.) Cyber Threat Intelligence, chap. 4, p. in press. Springer - Advances in Information Security series (2018)
2. Ding, Q., Li, Z., Haeri, S., Trajković, L.: Application of machine learning techniques to detecting anomalies in communication networks: Datasets and feature selection algorithms. In: M. Conti, A. Dehghantanha, T. Dargahi (eds.) Cyber Threat Intelligence, chap. 3, p. in press. Springer - Advances in Information Security series (2018)
3. Elingiusti, M., Aniello, L., Querzoni, L., Baldoni, R.: PDF-malware detection: a survey and taxonomy of current techniques. In: M. Conti, A. Dehghantanha, T. Dargahi (eds.) Cyber Threat Intelligence, chap. 9, p. in press. Springer - Advances in Information Security series (2018)

4. Haughey, H., Epiphaniou, G., Al-Khateeb, H., Dehghantanha, A.: Adaptive Traffic Fingerprinting for Darknet Threat Intelligence. In: M. Conti, A. Dehghantanha, T. Dargahi (eds.) Cyber Threat Intelligence, chap. 10, p. in press. Springer - Advances in Information Security series (2018)

5. Pandya, M.K., Homayoun, S., Dehghantanha, A.: Forensics Investigation of OpenFlow-Based SDN Platforms. In: M. Conti, A. Dehghantanha, T. Dargahi (eds.) Cyber Threat Intelligence, chap. 14, p. in press. Springer - Advances in Information Security series (2018)

6. Papalitsas, J., Rauti, S., Tammi, J., Leppänen, V.: A honeypot proxy framework for deceiving attackers with fabricated content. In: M. Conti, A. Dehghantanha, T. Dargahi (eds.) Cyber Threat Intelligence, chap. 12, p. in press. Springer - Advances in Information Security series (2018)

7. Park, R.: Guide to zero-day exploits (2015). URL `https://www.symantec.com/connect/blogs/guide-zero-day-exploits`

8. Petraityte, M., Dehghantanha, A., Epiphaniou, G.: A Model for Android and iOS Applications Risk Calculation: CVSS Analysis and Enhancement Using Case-Control Studies. In: M. Conti, A. Dehghantanha, T. Dargahi (eds.) Cyber Threat Intelligence, chap. 11, p. in press. Springer - Advances in Information Security series (2018)

9. Shackleford, D.: Who's using cyberthreat intelligence and how? – a SANS survey (2015). URL `https://www.sans.org/reading-room/whitepapers/analyst/cyberthreat-intelligence-how-35767`

10. Shalaginov, A., Banin, S., Dehghantanha, A., Franke, K.: Machine learning aided static malware analysis: A survey and tutorial. In: M. Conti, A. Dehghantanha, T. Dargahi (eds.) Cyber Threat Intelligence, chap. 2, p. in press. Springer - Advances in Information Security series (2018)

11. Wardman, B., Weideman, M., Burgis, J., Harris, N., Butler, B., Pratt, N.: A practical analysis of the rise in mobile phishing. In: M. Conti, A. Dehghantanha, T. Dargahi (eds.) Cyber Threat Intelligence, chap. 8, p. in press. Springer - Advances in Information Security series (2018)

12. Yasmin, R., Memarian, M.R., Hosseinzadeh, S., Conti, M., Leppänen, V.: Investigating the possibility of data leakage in time of live VM migration. In: M. Conti, A. Dehghantanha, T. Dargahi (eds.) Cyber Threat Intelligence, chap. 13, p. in press. Springer - Advances in Information Security series (2018)